\documentclass[journal]{IEEEtran}
\usepackage{tikz}
\usepackage{pgfplots,siunitx}
\pgfplotsset{compat=newest}
\pgfplotsset{width=7cm,compat=1.3}
\usepackage{graphicx}
\usepackage{algpseudocode}

\pgfplotsset{compat=newest}%
\usetikzlibrary{positioning,matrix,shapes.multipart,shapes.misc,spy}%
\definecolor{lines-1}{RGB}{228,26,28}
\definecolor{lines-2}{RGB}{55,126,184}
\definecolor{lines-3}{RGB}{77,175,74}
\definecolor{lines-4}{RGB}{152,78,163}
\definecolor{lines-5}{RGB}{255,127,0}
\definecolor{lines-6}{RGB}{153,153,153}
\definecolor{lines-7}{RGB}{166,86,40}
\definecolor{lines-8}{RGB}{247,129,191}
\definecolor{lines-9}{RGB}{255,255,51}
\pgfplotscreateplotcyclelist{myCycleList}{
	color=lines-1,semithick,mark=o,mark size=2,mark repeat=1\\%
	color=lines-2,semithick,mark=star,mark size=3,mark repeat=1\\%
	color=lines-3,semithick,mark=square,mark size=2,mark repeat=1\\%
	color=lines-4,semithick,mark=diamond,mark size=3,mark repeat=1\\%
	color=lines-5,semithick,mark=triangle,mark size=3,mark repeat=1\\%
	color=lines-6,semithick,mark=Mercedes star,mark size=3,mark repeat=1\\%
	color=lines-7,semithick,mark=|,mark size=3,mark repeat=1\\%
	color=lines-8,semithick,mark=x,mark size=3,mark repeat=1\\%
}
\pgfplotsset{
	compat=1.14,
	width =\columnwidth, 
	height=.8\columnwidth,
	ylabel absolute, ylabel style={yshift=-0.2cm},
	xlabel absolute, xlabel style={yshift=0.2cm},
	label style={font=\normalsize},
	tick label style={font=\scriptsize},
	legend style={font=\footnotesize,cells={align=left}},
	grid=both,
	minor grid style={dotted},
}

\usetikzlibrary{spy}
\usepackage[font=small]{caption}

\usepackage{balance}
\usepackage{multirow}
\usepackage{graphicx}
\usepackage{cite}
\usepackage{balance}
\usepackage{color}
\usepackage{subcaption}
\usepackage{diagbox} 
\usepackage{algorithm} 
\usepackage{amssymb}
\usepackage{amsfonts}
\usepackage[utf8]{inputenc} 
\usepackage[T1]{fontenc}
\usepackage{url}
\usepackage{ifthen}
\usepackage{cite}
\usepackage{gensymb}
\usepackage[cmex10]{amsmath} 

\newcommand{\R}{\mathbb{R}}
\newcommand{\Z}{\mathbb{Z}}
\newcommand{\B}{\mathcal{B}}

\newcommand{\Q}{\mathcal{Q}}
\newcommand{\X}{\mathcal{X}}
\newcommand{\I}{\mathcal{I}}

\newcommand{\ba}{\boldsymbol{a}}
\newcommand{\bc}{\boldsymbol{c}}
\newcommand{\bd}{\boldsymbol{d}}
\newcommand{\bl}{\boldsymbol{l}}
\newcommand{\bu}{\boldsymbol{u}}
\newcommand{\bv}{\boldsymbol{v}}
\newcommand{\bx}{\boldsymbol{x}}
\newcommand{\by}{\boldsymbol{y}}
\newcommand{\bz}{\boldsymbol{z}}

\newcommand{\bp}{\boldsymbol{p}}
\newcommand{\bG}{\boldsymbol{G}}
\newcommand{\bGs}{\boldsymbol{G}_\mathrm{s}}

\newcommand{\bI}{\boldsymbol{I}}
\newcommand{\bL}{\boldsymbol{L}}
\newcommand{\bU}{\boldsymbol{U}}
\newcommand{\bJ}{\boldsymbol{J}}
\newcommand{\bS}{\boldsymbol{S}}
\newcommand{\bT}{\boldsymbol{T}}
\newcommand{\bP}{\boldsymbol{P}}
\newcommand{\bR}{\boldsymbol{R}}
\newcommand{\bX}{\boldsymbol{X}}
\newcommand{\bY}{\boldsymbol{Y}}
\newcommand{\bzero}{\boldsymbol{0}}
\newcommand{\blambda}{\boldsymbol{\lambda}}

\newcommand{\Lambdas}{\Lambda_\mathrm{s}}

\DeclareMathOperator*{\argmin}{arg\,min} 

\ifCLASSINFOpdf
\else
\fi
\begin{document}
%
\title{Low-Complexity Voronoi Shaping \\for the Gaussian Channel}
%
%
%

\author{Shen~Li,
        Ali~Mirani,
        Magnus~Karlsson,~\IEEEmembership{Senior~Member,~IEEE,}~\IEEEmembership{Fellow,~OSA,}
        and~Erik~Agrell,~\IEEEmembership{Fellow,~IEEE}
\thanks{This work will be presented in part at the IEEE International Symposium on Information Theory (ISIT), Melbourne, Australia, July 2021.}
\thanks{This work was supported by the Swedish Research Council (VR) under grant no. 2017--03702.}
\thanks{S. Li and E. Agrell are with the Department
of Electrical Engineering, Chalmers University of Technology, 412 96 Gothenburg, Sweden. e-mail: shenl@chalmers.se.}
\thanks{A. Mirani and M. Karlsson are with the Department
of Microtechnology and Nanoscience, Chalmers University of Technology, 412 96 Gothenburg, Sweden.}
}

%
%

\markboth{Draft, \today}%
{Shell \MakeLowercase{\textit{et al.}}: Bare Demo of IEEEtran.cls for IEEE Journals}
%



\maketitle

\begin{abstract}
Voronoi constellations (VCs) are finite sets of vectors of a coding lattice enclosed by the translated Voronoi region of a shaping lattice, which is a sublattice of the coding lattice. In conventional VCs, the shaping lattice is a scaled-up version of the coding lattice. In this paper, we design low-complexity VCs with a cubic coding lattice of up to $32$ dimensions, in which pseudo-Gray labeling is applied to minimize the bit error rate. The designed VCs have considerable shaping gains of up to $1.03$ dB and finer choices of spectral efficiencies in practice. A mutual information estimation method and a log-likelihood approximation method based on importance sampling for very large constellations are proposed and applied to the designed VCs. With error-control coding, the proposed VCs can have higher achievable information rates than the conventional scaled VCs because of their inherently good pseudo-Gray labeling feature, with a lower decoding complexity.
\end{abstract}


%
\IEEEpeerreviewmaketitle

\section{Introduction}
%
%
%
%
\IEEEPARstart{P}{ower}
efficiency is important for higher-order modulation formats in communication systems. For an additive white Gaussian noise (AWGN) channel with an average power constraint, signal shaping is able to reduce the well-known asymptotic $1.53$ dB gap \cite{forney84} between the channel capacity and the achievable information rate (AIR) with a uniform signal (such as the most widely used quadrature amplitude modulation (QAM)), by adjusting the distribution of the transmitted signal alphabet to the capacity-achieving distribution, which is the Gaussian distribution. There are two flavors of shaping, geometric shaping by rearranging the positions of equally likely constellation points \cite{fengwensun93,forney84}, and probabilistic shaping by changing the distribution of a regular constellation \cite{calderbank90, kschischang93}. Much work has been devoted to the design of Gaussian-like constellations with feasible complexity in two \cite{gilbert52, campopiano62, foschini74, bin18,kadir20} and higher dimensions \cite{forney89a} for the AWGN channel.

The concept of Voronoi constellations (VCs) was first proposed in 1983 by Conway and Sloane, as a finite set of points of a lattice enclosed by a translated scaled-up version of its Voronoi region \cite{conway83}. Very fast algorithms for mapping integers to constellation points and vice versa were presented in \cite{conway83,conway82decoding}. In \cite{alijlt20,aliecoc}, Mirani \emph{et al.} designed multidimensional scaled VCs with up to $10^{28}$ constellation points utilizing this concept for the AWGN channel and nonlinear fiber channel, and showed high shaping gains and coding gains, and significant bit error rate (BER) and symbol error rate (SER) gains over QAM in uncoded system as well.

Forney generalized this concept in 1989 to VCs based on an arbitrary lattice partition $\Lambda/\Lambdas$ \cite{forney89b}, where $\Lambda$ is referred to as the coding lattice and $\Lambdas$ the shaping lattice. They can be different, as long as the shaping lattice is a sublattice of the coding lattice. Forney presented encoding and decoding algorithms for certain choices of the shaping lattice, such as so-called mod-2 or mod-4 binary lattices.

Feng \emph{et al.} presented a more general method to enumerate the points in a VC based on an arbitrary lattice partition \cite{feng13}, which is reviewed by Zamir in \cite[Ch.~9]{zamir14book}, \cite{zamir14}. This enumeration admits encoding and decoding algorithms as fast as Conway and Sloane's. An equally simple but less general method called ``rectangular encoding'' was proposed by Kurkoski in \cite{kurkoski18} for VCs whose shaping lattice and coding lattice both have triangular generator matrices. This method is applicable to a variety of coding lattices and shaping lattices.

Ferdinand \emph{et al.} proposed a two-step VC construction method called ``systematic Voronoi shaping'' in \cite{ferdinandTWC} based on the concept of ``systematic shaping'' proposed by Sommer \emph{et al.} in \cite{sommer09}, combining a high-dimensional coding lattice defined by a lower-triangular parity check matrix and a lower-dimensional shaping lattice, to achieve high coding gains and high shaping gains.  The SER performance was evaluated when low-density lattice codes \cite{sommer08} are used as the coding lattice and some common multidimensional lattices with low-complexity quantization algorithms are used for the shaping lattice. For the shaping step, algorithms to map the integers to points in VCs with a cubic coding lattice and vice versa were explicitly described in \cite{ferdinand14}.


In our conference paper \cite{ourISIT}, we studied VCs with a cubic coding lattice, for which we compared Feng's, Ferdinand's, and Kurkoski's encoding and decoding algorithms. To minimize the BER, we applied pseudo-Gray labeling to these algorithms, and evaluated the performance of the designed VCs in terms of the Gray penalty \cite{simon1973,smith1975} and BER performance for some common multidimensional shaping lattices both in uncoded and coded systems. In our proposed scheme, coding is completely separated from shaping and performed using error-control coding, outside the Voronoi shaping.

We study VCs with a cubic coding lattice for the following three reasons. First, the decoding algorithm is much simpler than for VCs with rescaled coding and shaping lattices \cite{conway83, alijlt20}, since the search for the closest lattice point, which dominates the decoding complexity for high-dimensional lattices, is simple dimension-wise integer rounding for a cubic coding lattice. Second, although the proposed VCs have no coding gain, the shaping gain is still achievable, and the lack of coding gain can be compensated by error-correction coding that is usually performed anyway. After error-correction coding, the proposed VCs can have better BER performance than the conventional scaled VCs because of their inherently good pseudo-Gray labeling feature. Third, for mapping integers to bits, the proposed VCs allow for improved granularity in spectral efficiencies.

 In this paper, we extend our work \cite{ourISIT} on the VCs with a cubic coding lattice. We propose a new method based on the concept of importance sampling to estimate the mutual information (MI) for very large constellations, and exemplify it for VCs. This has been considered as a challenging issue for such large constellations since the exact MI calculation requires enumerating all constellation points. The MI of VCs is evaluated here for the first time to our knowledge. The AIR performance of VCs in combination with a low-density parity-check (LDPC) code is also presented. For mapping integers to bits, we improve the granularity in spectral efficiencies by rotating the shaping lattice by $\ang{45}$ and scaling it by $\sqrt{2}$. We also investigate the trade-off between shaping gain and decoding complexity in higher dimensions.
 

\emph{Notation:} Bold lowercase symbols denote row vectors and bold uppercase symbols denote matrices or random vectors. The elements of a vector $\bu$ are denoted by $u_i$, the rows of a matrix $\bP$ are denoted by $\bp_i$, and the element at row $i$, column $j$ of a matrix $\bP$ are denoted by $P_{ij}$. The sets of integer, real, complex, and natural numbers are denoted by $\Z$, $\R$, $\mathbb{C}$, and $\mathbb{N}$, respectively. Rounding a vector to its nearest integer vector is denoted by $\lfloor \cdot \rceil$, in which ties are broken arbitrarily. The largest integer not greater than a given real number is denoted by $\lfloor \cdot \rfloor$.

\section{Preliminaries}

Given a set of $n$ linearly independent \emph{basis vectors,} a \emph{lattice} is the set of all linear combinations of these vectors with integer coefficients. If the basis vectors are arranged row-wise into a matrix $\bG$, then the lattice is
\begin{align} \label{eq:latticedef}
\Lambda \triangleq \{ \bu \bG :\; \bu \in \Z^n \}
.\end{align}
Without loss of generality, we assume that the generator matrix has dimension $n \times n$.%
\footnote{The $n$ basis vectors must have dimension at least $n$ in order to be linearly independent. If $n$ basis vectors are given in more than $n$ dimensions, then an $n$-dimensional lattice, which is equivalent to the original lattice in Euclidean geometry, can be defined by rotating (e.g., QR-decomposing) $\bG$.} From the definition, any lattice includes the all-zero vector $\bzero$. The generator matrix of a given lattice is not unique. Two generator matrices $\bG$ and $\bG'$ generate the same lattice if and only if $\bG' = \bU \bG$, where $\bU$ is an integer matrix with determinant $\pm 1$ \cite[p.~10]{conway99book}.

The \emph{closest lattice point quantizer} $\Q_{\Lambda}(\cdot)$ maps an arbitrary vector $\bx\in\R^n$ to its closest lattice point in $\Lambda$
\begin{align}
    \Q_{\Lambda}(\bx)=\argmin_{\blambda \in \Lambda}\|\bx-\blambda\|^2.
\end{align}

If an affine transformation function $T(\cdot)$ is applied to the lattice $\Lambda$, e.g., a scaling, rotation, or/and a shift, then it follows that
\begin{align}
    \mathcal{Q}_{T(\Lambda)}(\bx )=T(\mathcal{Q}_{\Lambda}(T^{-1}(\bx))).\label{eq:T}
\end{align}

The \emph{fundamental Voronoi region} of a lattice $\Lambda$ is the set of vectors in Euclidean space having the all-zero vector as its closest lattice point, i.e.,
\begin{align}
\Omega(\Lambda) \triangleq \{\bx \in \R^n :\; \Q_{\Lambda}(\bx)=\bzero \}
.\end{align}
Given an $n$-dimensional coding lattice $\Lambda$, an $n$-dimensional shaping lattice $\Lambdas$, and an \emph{offset vector} $\ba \in \R^n$, where $\Lambdas$ is a sublattice of $\Lambda$, i.e., $\Lambdas\subset \Lambda$, a VC in its general form defined by Forney \cite{forney89b} is
\begin{align}
\Gamma \triangleq (\Lambda-\ba) \cap \Omega(\Lambdas)\label{eq:VC}
.\end{align}
We assume that no points in $\Lambda-\ba$ fall on the boundary of $\Omega(\Lambdas)$.%
\footnote{An arbitrarily small perturbation of the offset $\ba$ in a random direction prevents points from falling on the boundary with probability one.} The number of points in the VC is
\begin{align}
M \triangleq |\Gamma| = \frac{|\!\det\bGs|}{|\!\det\bG|}
,\end{align}
where $\bGs$ is a generator matrix of $\Lambdas$. This relation can be verified by recognizing $|\!\det\bG|$ and $|\!\det\bGs|$ as the volumes of $\Omega(\Lambda)$ and $\Omega(\Lambdas)$, respectively \cite[p.~4]{conway99book}. The average symbol energy is
\begin{align}
E_{\text{s}}=\frac{1}{M}\sum_{\bx\in\Gamma}\|\bx\|^{2}.\label{eq:Es}
\end{align}

To compare the performance of different constellations, the following relevant figures of merit were defined and widely used in the literature.

\emph{1)} The \emph{spectral efficiency} \cite{forney89a, kschischang93, mkeaofc12} of a constellation is defined as
\begin{align}
    \beta=2\log_{2}(M)/n ~\text{[bits/symbol/dimension-pair]}.
\end{align}

\emph{2)} The \emph{asymptotic power efficiency} (APE) \cite[Eq.~(5.8)]{Benedetto99book},\cite{agrell09} is defined as
\begin{align}
\gamma=\frac{d_{\text{min}}^{2}\log_{2}(M)}{4E_{\text{s}}},
\end{align}
where $d_{\text{min}}$ is the minimum Euclidean distance of the constellation. Usually, the pulse amplitude modulation (PAM) is chosen as the benchmark \cite{forney89a,kschischang93}, which has an APE of 
\begin{align}
\gamma_{\text{PAM}}=\frac{3\beta}{2(2^{\beta}-1)}.
\end{align}
This also applies to the geometric extension of a PAM, i.e., an $n$-dimensional cubic constellation constructed by the Cartesian product of $n$ equal one-dimensional PAM constellations.

\emph{3)} The \emph{APE gain} of a constellation over a cubic constellation at the same spectral efficiencies is quantified as
\begin{align}
    g=10\log_{10}\frac{\gamma}{\gamma_{\text{PAM}}}~\text{[dB]}.
\end{align}
It can be divided into a \emph{coding gain} $g_{\text{c}}$ obtained by packing the constellation points in a cubic shape more densely, and a \emph{shaping gain} $g_{\text{s}}$ obtained by making the boundary of the cubic-packing points more spherical \cite{forney89a,mkeaofc12}. For the fundamental Voronoi region of the lattice $\Lambdas$, the (asymptotic) shaping gain is defined as $g_{\text{s}}(\Lambdas)=1/(12G(\Omega(\Lambdas)))$ \cite{forney89a}, where $G(\Omega(\Lambdas))$ is the normalized second moment of this region \cite[Eq.~(9)]{conway82Gvalue}.


\section{Mapping integers to VCs}
 Feng \emph{et al.} proposed encoding and decoding algorithms, i.e., mapping integers to constellation points and vice versa, for arbitrary VCs in \cite{feng13}, which relies on the Smith normal form \cite[Ch.~15]{brownbook93}. In \cite{ferdinand14}, Ferdinand \emph{et al.} proposed encoding and decoding algorithms specifically for VCs with a cubic coding lattice. Kurkoski proposed a ``rectangular encoding'' method in \cite{kurkoski18}, which is applicable to VCs with a shaping lattice described by a triangular generator matrix and a cubic coding lattice. In this section, the three algorithms are reviewed and compared specifically when they are applied to VCs with a cubic coding lattice.

Table \ref{tab:algorithms} summarizes the three algorithms for the special case of the lattice partition $\Z^n/\Lambdas$, which is the focus of this paper. The more general versions of Feng's algorithms for arbitrary $\Lambda$ are described in \cite{feng13}. In Ferdinand's algorithms described in \cite{ferdinand14}, the basis vectors are column vectors and the decoding goes from low to high dimensions. Here for the convenience of comparison, we describe them using row vectors as basis vectors, and rewrite the decoding process of \cite[Eqs.~(19)--(21)]{ferdinand14}, with a reverse order of decoding, which however does not change the performance of their algorithms. 

The closest lattice point quantizer $\Q_{\Lambdas}(\cdot)$ in these algorithms is well-studied for many common lattices \cite[Ch.~20]{conway99book}, \cite{conway82decoding, conway84}, and other lattices can be handled by general algorithms \cite{agrell02,Ghasemmehdi11}.

\begin{table*}[tbp]
  \newcommand{\tabincell}[2]{\begin{tabular}{@{}#1@{}}#2\end{tabular}}
  \centering
  \begin{tabular}{| m{3cm} | m{4.75cm}| m{4.1cm} | m{4.1cm}|}
    \hline
    Algorithms & Feng's algorithms\cite{feng13}& Ferdinand's algorithms \cite{ferdinand14}& Kurkoski's algorithms\cite{kurkoski18}\\
    \hline
    \hline
    \emph{Preprocessing:} Given the generator matrix $\bGs$ of the shaping lattice $\Lambdas$ and the generator matrix of the cubic coding lattice is $\bG=\bI_{n}$. 
    &Find two integer matrices $\bS$ and $\bT$ with determinant $\pm 1$ such that $\bJ=\bS\bGs\bT$ is the Smith normal form of $\bGs$. Then let $u_i \in \{0,\ldots, J_{ii}-1\}$ for $i=1,\ldots,n$.
    &Find an integer matrix $\bS$ with determinant $\pm 1$ such that $\bL=\bS\bGs$ is a lower-triangular matrix with positive diagonal elements. Let $u_i \in \{0,\ldots, L_{ii}-1\}$ for $i=1,\ldots,n$.
    &Find an integer matrix $\bS$ with determinant $\pm 1$ such that $\bL=\bS\bGs$ is a lower-triangular matrix with positive diagonal elements. Let $u_i \in \{0,\ldots, L_{ii}-1\}$ for $i=1,\ldots,n$. \\
    \hline
    \emph{Encoding:} Input $\bu$. Output $\bx$.
    &\tabincell{l}{1) Let $\bc \leftarrow \bu \bT^{-1} - \ba$\\2) Let $\bz \leftarrow \Q_{\Lambdas}(\bc)$\\3) Let $\bx \leftarrow \bc-\bz$}
    & \tabincell{l}{1) Let $\bd \leftarrow \left ( \frac{u_{1}}{L_{11}},\frac{u_{2}}{L_{22}},\ldots,\frac{u_{n}}{L_{nn}} \right )$\\2) Let $\bc=\bd\bL-\ba$\\3) Let $\bz \leftarrow \Q_{\Lambdas}(\bc)$\\4) Let $\bx \leftarrow \bc-\bz$}
    & \tabincell{l}{1) Let $\bc \leftarrow \bu - \ba$\\2) Let $\bz \leftarrow \Q_{\Lambdas}(\bc)$\\3) Let $\bx \leftarrow \bc-\bz$}\\
    \hline
    \emph{Decoding:} Input $\by$. Output $\bu$.
    &\tabincell{l}{1) Let $\bc \leftarrow \lfloor \by+\ba \rceil$\\2) Let $\bu \leftarrow \bc \bT$\\3) Let $u_i \leftarrow u_i \mod J_{ii} $, $\forall i=1,\ldots,n$}
    & \tabincell{l}{1) Let $\bu \leftarrow \lfloor \by+\ba \rceil$\\2) For $i=n,n-1,\ldots,1,$ do\\~~~~~$\left\{\begin{matrix}
v_i \leftarrow \lfloor u_i/L_{ii} \rfloor\\ \bu \leftarrow \bu-v_i \bl_i \end{matrix}\right.$} 
    &\tabincell{l}{1) Let $\bu \leftarrow \lfloor \by+\ba \rceil$\\2) for $i=n,n-1,\ldots,1,$ do\\~~~~~$\left\{\begin{matrix}
v_i \leftarrow \lfloor u_i/L_{ii} \rfloor\\ \bu \leftarrow \bu-v_i \bl_i \end{matrix}\right.$}\\
    \hline
  \end{tabular}
  \caption{A Comparison of different encoding and decoding algorithms for VCs with a cubic coding lattice.}
  \label{tab:algorithms}
\end{table*}

Every matrix over a principal ideal domain has a Smith normal form, and the integers form a principal ideal domain \cite{brownbook93,feng13}. In Feng's algorithm, given an integer generator matrix $\bGs$, the Smith normal form is first computed in the preprocessing stage as $\bJ=\bS\bGs\bT$, where $\bJ$ is a diagonal matrix with positive diagonal elements and $J_{ii}$ divides $J_{ii+1}$ for $i=1,\ldots,n-1.$ Then the integer vectors $\bu$ are enumerated according to the diagonal elements of $\bJ$. The $M=\!\det \bJ$ possible values of $\bu\bT^{-1}$ can be uniquely mapped to the $M$ constellation points $\bx\in\Gamma$ by
\begin{align} \label{eq:enumerationFeng}
\bx = \bu \bT^{-1} + \bv \bS\bGs - \ba,
\end{align}
where $\bv\in\Z^n$. The Smith normal form $\bJ$ is unique given a $\bGs$, whereas $\bS$ and $\bT$ are not unique; thus Feng's algorithms can generate different mapping rules for the same $\bGs$.

In Ferdinand's algorithms, a lower-triangular generator matrix $\bL$ for $\Lambdas$ is first computed (the algorithms are applicable only if $\bL$ exists), then the integer vectors $\bu$ are enumerated according to the diagonal elements of $\bL$. In the encoding process, $\bd\bL$ are defined to be integer vectors in the fundamental parallelotope of $\Lambdas$. The constellation points $\bx \in \Gamma$ can be labeled by unique values of $\bd\bL$ and $\bv\in \Z^n$ as
\begin{align} \label{eq:enumerationFerdinand}
\bx = \bd \bL + \bv \bL - \ba.
\end{align}
 However, $\bd\bL$ are not guaranteed to be integer vectors for arbitrary shaping lattices, thus making Ferdinand's algorithms less general than Kurkoski's algorithms. The condition for Ferdinand's algorithms to be applicable is 
\begin{align}
    \frac{L_{ij}}{L_{ii}}\in \Z, \forall i={1,\ldots,n}, ~1\leq j<i.
\end{align}
The decoding can be done sequentially, beginning from $u_n$, thanks to the triangular structure of $\bL$. 

Kurkoski's algorithms are also applicable only if the lower-triangular matrix $\bL$ exists. Then every point $\bx\in\Gamma$ can be uniquely enumerated by vectors $\bu$ and $\bv$ such that
\begin{align} \label{eq:enumeration}
\bx = \bu + \bv \bGs - \ba,
\end{align}
where $u_i \in \{0,\ldots, L_{ii}-1\}$ for $i=1,\ldots,n$ and $\bv \in \Z^n$. There are $M=\!\det \bL = \prod_i L_{ii}$ possible values of $\bu$, and each of them occurs exactly once among all points $\bx\in\Gamma$. The decoding process is the same as Ferdinand's.

Given the same $\bGs$, Feng's and Ferdinand's algorithms can have different mapping rules, and Kurkoski's algorithms provide a third mapping rule, see the numerical example in this section. When Feng's $\bu\bT^{-1}$ and Ferdinand's $\bd\bL$ are equal, their algorithms are equivalent. For most commonly used shaping lattices, e.g., $D_{4}$, $E_{8}$, $\Lambda_{16}$, and $\Lambda_{24}$ \cite[Ch.~4]{conway99book}, Ferdinand's algorithms are applicable. Feng's algorithms can have the same mapping rule as Ferdinand's algorithms, when the generator matrices are written in nice lower-triangular matrices as in \cite[Ch.~4]{conway99book}. 

Apparently Feng's algorithms are more general than Kurkoski's, which are in turn slightly more general than Ferdinand's. However, for the case where all three algorithms are applicable, e.g., for commonly used multidimensional shaping lattices, Kurkoski's algorithms generally provide better labelings of constellation points. When a Gray code is applied to label the integer coordinates of $\bu$, i.e., converting $u_{i}$ to a binary reflected Gray code for $i=1,\ldots,n$, each pair of nearest $\bu$ in terms of Euclidean distance differs by exactly one bit. Kurkoski's algorithms allow for directly mapping $\bu$ to constellation points, thus making most of the constellation points have Gray neighbors, which we call pseudo-Gray labeling. In Feng's and Ferdinand's encoding algorithms, $\bu$ is multiplied by a matrix, which changes the neighbor relationships between constellation points. Kurkoski's algorithms allow for this pseudo-Gray labeling for VCs, and can thus achieve a smaller Gray penalty and better BER performance \cite{ourISIT} than Feng's and Ferdinand's algorithms. Throughout this paper, we adopt Kurkoski's encoding and decoding and this pseudo-Gray labeling scheme.  

\begin{figure*}[tbp]
\centering
\subfloat[Feng's algorithms.]{\includegraphics[width=6.6in]{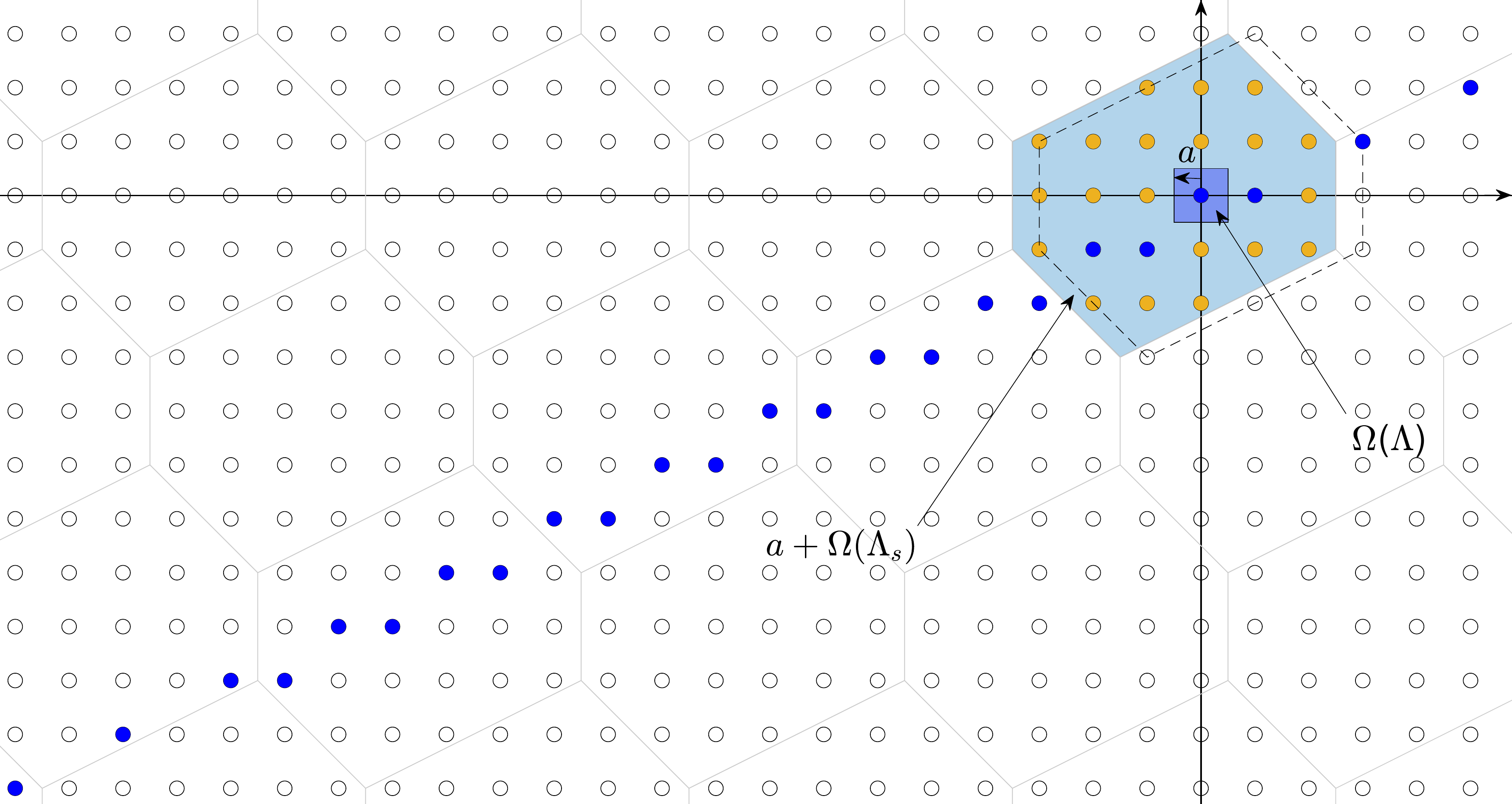}%
\label{Fig:FengVC}}
\hfil
\subfloat[Ferdinand's algorithms.]{\includegraphics[width=3.1in]{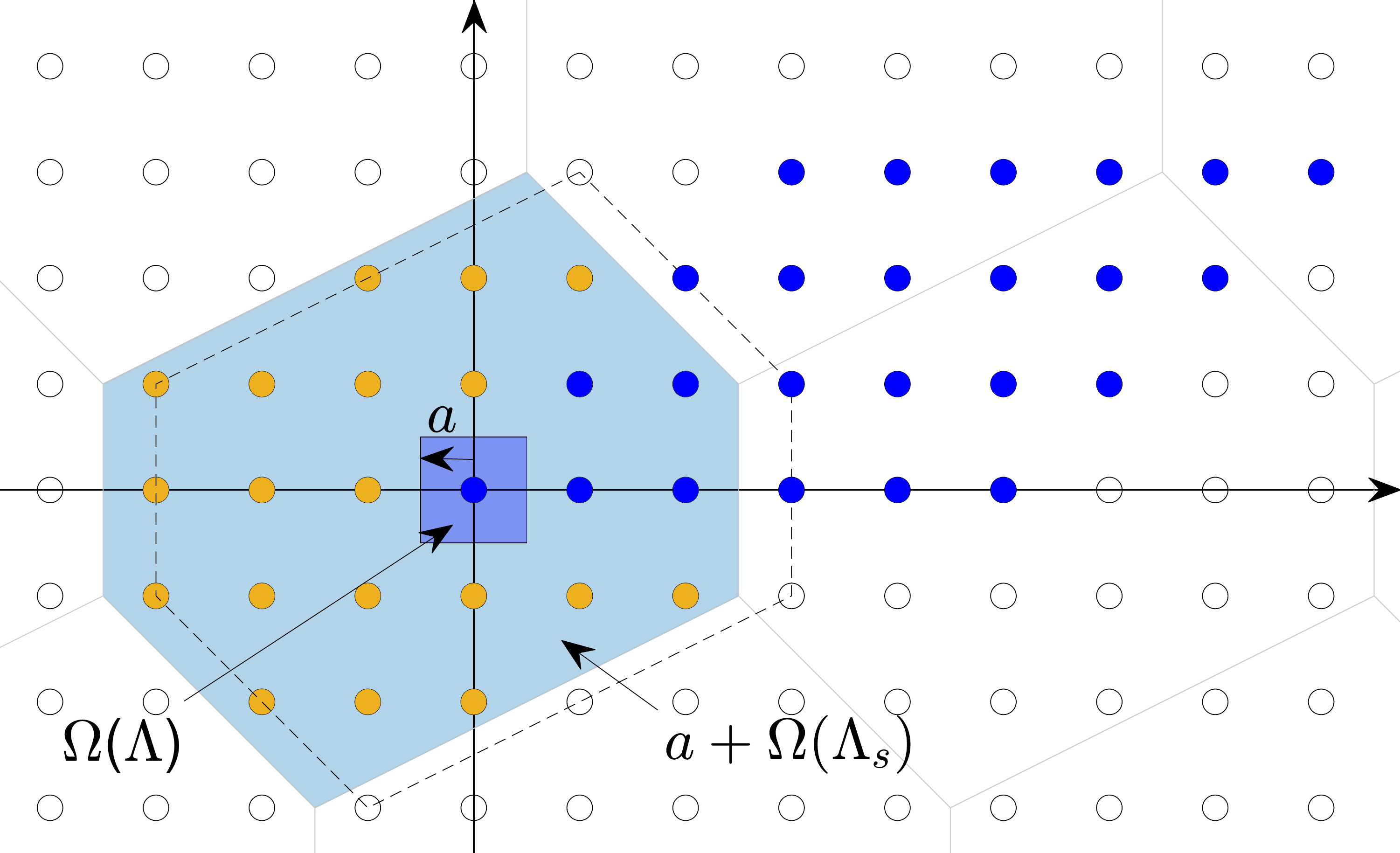}%
\label{Fig:FerdinandVC}}
\hspace{0.4in}
\subfloat[Kurkoski's algorithms.]{\includegraphics[width=3.1in]{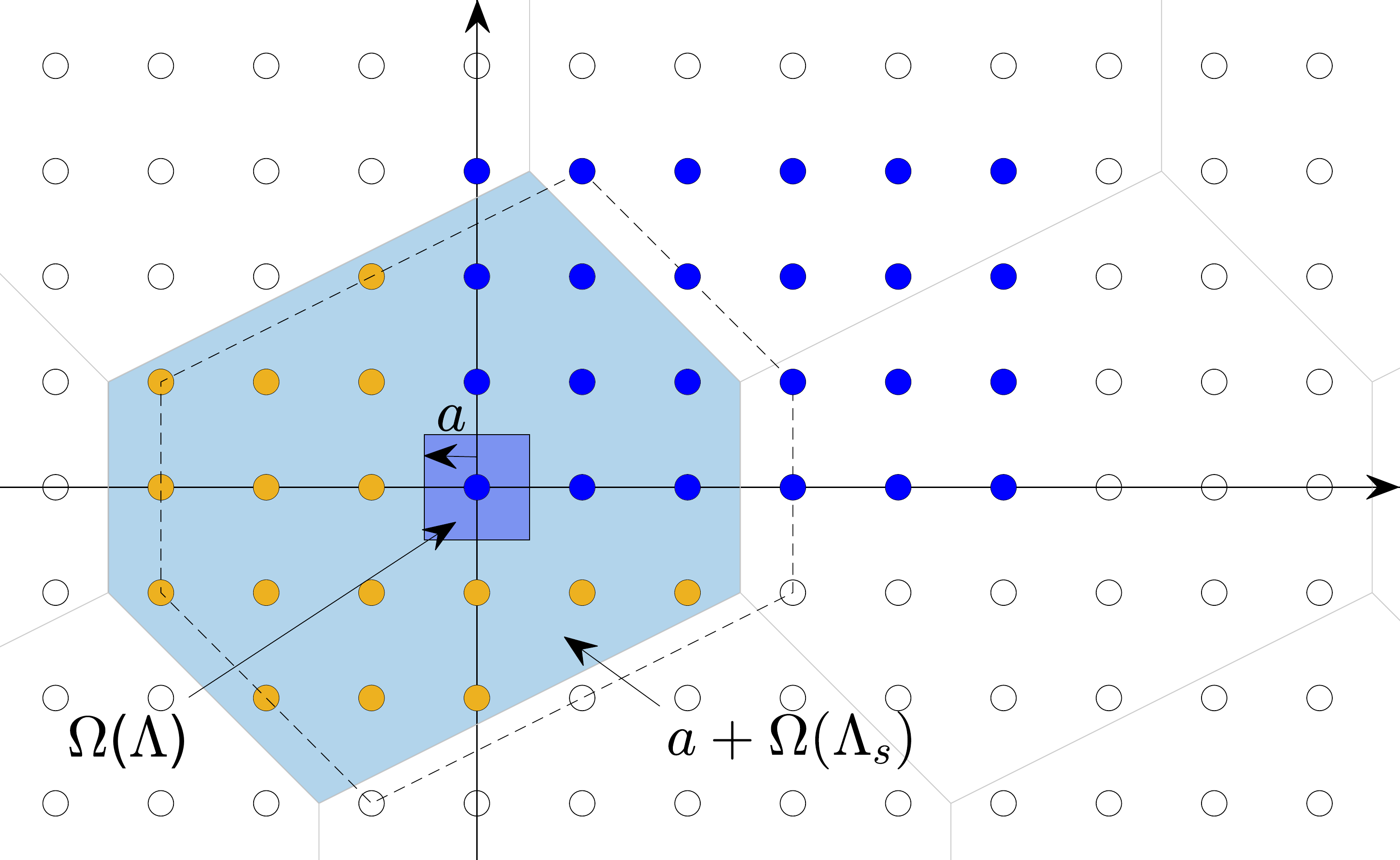}%
\label{Fig:ourVC}}
\caption{Example: different integer mapping rules for a two-dimensional VC. The blue filled points are encoded into points in the shifted Voronoi region $\ba+\Omega(\Lambdas)$ (the light blue region) in encoding.}
\label{Fig:VC}
\end{figure*}

\emph{Example:} We consider a two-dimensional VC for which the generator matrices of $\Lambdas$ and $\Lambda$ are
\begin{align}
\bGs=\begin{pmatrix}
6 & 0\\ 
4 & 4
\end{pmatrix},\;
\bG=\begin{pmatrix}
1 & 0\\ 
0 & 1
\end{pmatrix},
\end{align}
and the offset vector $\ba=(-1/2, 0)$.

In Feng's algorithms, the Smith normal form $\bJ=\bS \bGs \bT$ is
\begin{align}
    \bJ=\begin{pmatrix}
    2&0\\
    0&12
    \end{pmatrix},\;
    \bS=\begin{pmatrix}
    1&1\\
    -2&-3
    \end{pmatrix},\;
    \bT=\begin{pmatrix}
    1&2\\
    -2&-5
    \end{pmatrix}.
\end{align}
Then the integer vectors $\bu=(u_1,u_2)$ are defined as ${u_1 \in \{0,1\}}$ and $u_2 \in \{0,\ldots, 11\}$. The vectors $\bu\bT^{-1}$ are calculated as
\begin{align}
\bu\bT^{-1}=
    \begin{pmatrix}
    u_{1}& u_{2}
    \end{pmatrix}
    \begin{pmatrix}
    5& 2\\
    -2& -1
    \end{pmatrix}
=\begin{pmatrix}
    5u_{1}-2u_{2}&2u_{1}-u_{2}
    \end{pmatrix}.
\end{align}

In Ferdinand's algorithms, $\bL=\bGs$ and $\bu$ are enumerated as $u_1 \in \{0,\ldots, 5\}$ and $u_2 \in \{0,\ldots, 3\}$. The vectors $\bd\bL$ are computed by
\begin{align}
\bd\bL=
    \begin{pmatrix}
    u_{1}/6& u_{2}/4
    \end{pmatrix}
    \begin{pmatrix}
    6& 0\\
    4& 4
    \end{pmatrix}
=\begin{pmatrix}
    u_{1}+u_{2}&u_{2}
    \end{pmatrix}.
\end{align}
In Kurkoski's algorithms, $\bL=\bGs$ and $\bu$ are enumerated as $u_1 \in \{0,\ldots, 5\}$ and $u_2 \in \{0,\ldots, 3\}$.

Fig.~\ref{Fig:VC} illustrates the three different mapping rules for this example, where the vectors $\bu\bT^{-1}$ in Feng's algorithms, $\bd\bL$ in Ferdinand's algorithms, and $\bu$ in Kurkoski's algorithms are highlighted. The enumeration of points in a rectangular shape of Kurkoski's algorithms help achieve a lower Gray penalty and a lower BER \cite{ourISIT}.

\section{Design of VCs}
A VC based on the lattice partition $\Lambda/\Lambdas$ is defined by \eqref{eq:VC}. In this section, we discuss the choice of parameters of VCs based on the lattice partition $\Z^n/\Lambdas$.

 To minimize the average symbol energy defined by \eqref{eq:Es}, the offset vector $\ba$ is optimized using an iterative algorithm given in \cite{conway83}. This algorithm in may converge to a suboptimal vector for $8$ and higher dimensions, and cannot be exactly calculated when $M$ is very large. However, as $M$ increases, the performance difference between VCs generated using the optimal $\ba$ and a random $\ba \in \Omega(\Lambda)$ decreases, and can be neglected for large VCs, see \cite[Fig.~3]{alijlt20}. In this paper, for a small or moderate-size VC ($M\leq 2^{17}\approx 1.3\times10^5$), $\ba$ was optimized using the method in \cite{conway83}, whereas for very large VCs where we can only approximate the average symbol energy by Monte Carlo simulations, a random $\ba$ uniformly distributed in $\Omega(\Lambda)$ was selected.

The shaping lattice $\Lambdas$ should have a high shaping gain and low-complexity quantization algorithm. The most commonly-used multidimensional shaping lattices: the $4$-dimensional checkerboard lattice $D_{4}$, $8$-dimensional lattice $E_{8}$, $16$-dimensional Barnes--Wall lattice $\Lambda_{16}$, the $24$-dimensional Leech lattice $\Lambda_{24}$ \cite[Ch.~4]{conway99book}, and a suboptimal $32$-dimensional lattice $L_{32}$ are considered in this paper. The $32$-dimensional lattice $L_{32}$ is constructed by applying Construction B \cite[Ch.~5]{conway99book} to the $(32,6,16)$ first order Reed--Muller code. This lattice has a sublattice $2D_{32}$ of index $64$, which is beneficial, since the closest lattice point quantizer dominates the complexity for the encoding and decoding, especially for high-dimensional lattices. Compared with $\Lambda_{24}$ which has a sublattice $4D_{24}$ of index $8192$ \cite{conway84}, the complexity is reduced by a factor of $128$, while the shaping gain is only $0.091$ dB less. Thus, $L_{32}$ can be considered as a good trade-off between the shaping gain and decoding complexity. Table \ref{tab:gammas} shows the normalized second moment $G(\Omega(\Lambdas))$ and shaping gains $g_{s}(\Lambdas)$ for these considered shaping lattices.

\begin{table}
  \renewcommand{\arraystretch}{1.3}
  \caption{Asymptotic shaping gains and normalized second moments of Voronoi regions of some lattices.}
  \label{tab:gammas}
  \centering
  \begin{tabular}{c c c c c c}
    \hline 
    $\Lambdas$ &$D_{4}$& $E_{8}$&$\Lambda_{16}$& $\Lambda_{24}$ &$L_{32}$\\
    \hline \hline
    $G(\Omega(\Lambdas))$ & $0.0766$ & $0.0717$ & $0.0683$ & $0.0658$ & $0.0671$\\
    \hline
    $g_{\text{s}}(\Lambdas)$ [dB] & $0.366$ & $0.653$ & $0.864$ & $1.026$& $0.935$\\
    \hline
  \end{tabular}
\end{table}
A VC based on the lattice partition $\Lambda/\Lambdas$ with ${M=|\!\det\bGs|/|\!\det \bG|}$ constellation points can be scaled by any integer, i.e., $\Lambda/m\Lambdas,\;m\in \Z$. The new VC after scaling has a spectral efficiency of 
\begin{align}
    \beta&=\frac{2\log_{2}(|\!\det m^n\bGs|/|\!\det\bG|)}{n} \notag\\
    &=\frac{2\log_{2}(M)}{n}+2\log_{2}(m).\label{eq:beta}
\end{align}

A rotation of $\ang{45}$ together with a scaling of $\sqrt{2}$ of the shaping lattice $\Lambdas$ can provide $1$ more bit/symbol/dimension-pair. This can be done for an even $n$ by multiplying the generator matrix $\bGs$ from the right side with an $n$-by-$n$ matrix \cite{forney88}
\begin{align}
\bR=\begin{pmatrix}
 1& 1 & 0 & 0 &0  & \cdots & 0 &0 \\ 
 -1&  1&  0&  0&  0&  \cdots&  0&0 \\ 
 0& 0 & 1 & 1 &  0&  \cdots&  0&0 \\ 
 0& 0& -1& 1&  0& \cdots & 0 &0 \\ 
 0& 0&  0&  0&  \ddots& ~&  \vdots& \vdots\\ 
 \vdots& \vdots&  \vdots&  \vdots& ~& \ddots &  0& 0\\ 
 0& 0&  0&  0&  \cdots&  0& 1 & 1\\ 
 0& 0&  0&  0&  \cdots&  0& -1& 1
\end{pmatrix},
\end{align}
which operates on every dimension-pair of $\Lambdas$, with a determinant of $2^{n/2}$. The generator matrix of the new shaping lattice $m\Lambdas \bR$ (which remains a sublattice of $\Lambda$) is $m\bGs\bR$, and the closest point quantizer becomes
\begin{align}
    \Q_{m\Lambdas\bR}(\bx)=m\Q_{\Lambdas}(\frac{1}{m}\bx\bR^{-1})\bR
\end{align}
by \eqref{eq:T}. The new VC based on the lattice partition $\Lambda/m\Lambdas\bR$ has a spectral efficiency of
\begin{align}
    \beta&=\frac{2\log_{2}(|\!\det m^n\bGs\bR|/|\!\det\bG|)}{n} \notag\\
    &=\frac{2\log_{2}(M)}{n}+2\log_{2}(m)+1.\label{eq:beta_rotation}
\end{align}

\begin{figure}[tbp]
\centering
\begin{tikzpicture}
	\begin{axis}[
		xmin=0,
		xmax=13.5,
		ymin=0.2, ymax=1.2,
		xlabel={$\beta$ [bits/symbol/dimension-pair]},
		ylabel={APE gain $g$ [dB]},
		ylabel style={at={(axis description cs:-0.01,0.5)}, anchor=north},
		cycle list name=myCycleList,
	    legend pos=north west,
		legend cell align=left,
		legend style={fill=white,fill opacity=0.8, draw opacity=1,text opacity=1},
		ylabel style={yshift=-.05cm},
		xlabel style={xshift=-.05cm},
		height =0.4\textwidth,
		width=0.5\textwidth,
	]

	\addplot+[lines-1] table[
		x=SE,
		y=gammas,
	] {./figures/SE/gammasD4.txt};\addlegendentry{$\Z^{4}/D_{4}$}
 	\addplot+[lines-2] table[
 		x=SE,
 		y=gammas,
 	] {./figures/SE/gammasE8.txt};\addlegendentry{$\Z^{8}/E_{8}$}
 	\addplot+[lines-3] table[
		x=SE,
 		y=gammas,
 	] {./figures/SE/gammasL16.txt};\addlegendentry{$\Z^{16}/\Lambda_{16}$}
 	 \addplot+[lines-4] table[
		x=SE,
 		y=gammas,
 	] {./figures/SE/gammasL24.txt};\addlegendentry{$\Z^{24}/\Lambda_{24}$}
 	\addplot+[lines-5] table[
		x=SE,
 		y=gammas,
 	] {./figures/SE/gammasL32.txt};\addlegendentry{$\Z^{32}/L_{32}$}

 	 \addplot+[dashed,color=black,mark=none] coordinates {(-2,0.3659) (19,0.3659)};\addlegendentry{$g_{\text{s}}(\Lambdas)$}
 	\addplot+[only marks,lines-1,mark=o,mark size=1] table[
		x=SE,
		y=gammas,
	] {./figures/SE/gammas_scaled_D4.txt};
	\addplot+[only marks, lines-2,mark=star,mark size=1] table[
		x=SE,
		y=gammas,
	] {./figures/SE/gammas_scaled_E8.txt};
	\addplot+[only marks, lines-3,mark=square,mark size=1] table[
		x=SE,
		y=gammas,
	] {./figures/SE/gammas_scaled_L16.txt};
	\addplot+[only marks, lines-4,mark=diamond,mark size=1] table[
		x=SE,
		y=gammas,
	] {./figures/SE/gammas_scaled_L24.txt};
	\addplot+[only marks, lines-5,mark=triangle,mark size=1] table[
		x=SE,
		y=gammas,
	] {./figures/SE/gammas_scaled_L32.txt};
 	 \addplot+[dashed,color=black,mark=none] coordinates {(-2,0.6530) (19,0.6530)};
 	 \addplot+[dashed,color=black,mark=none] coordinates {(-2,0.8640) (19,0.8640)};
 	 \addplot+[dashed,color=black,mark=none] coordinates {(-2,1.0259) (19,1.0259)};
 	 \addplot+[dashed,color=black,mark=none] coordinates {(-2,0.9345) (19,0.9345)};
	\end{axis}
\end{tikzpicture}
\caption{The APE gain $g$ as a function of $\beta$ for VCs with a cubic coding lattice. The smaller markers on top of the lines represent the cases in which the scaling factor is not a power of $2$. The black dashed lines are the asymptotic shaping gains $g_{\text{s}}(\Lambdas)$ for these shaping lattices stated in Table \ref{tab:gammas}.}
\label{Fig:shapinggain}
\end{figure}
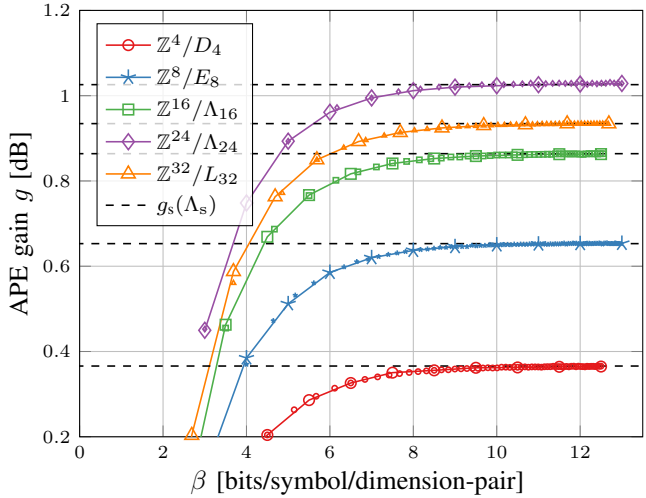

In order to make bit mapping possible, both $M$ and $m$ must be powers of $2$. With $m$ an arbitrary power of $2$, $\Lambda/m \Lambda_s$ gives a resolution of $2$ bits/symbol/dimension-pair according to \eqref{eq:beta}. However, $\Lambda/m \Lambda_s \bR$ offers the intermediate spectral efficiencies as in \eqref{eq:beta_rotation}, which decreases the overall resolution to $1$ bit/symbol/dimension-pair. Also, the combination of different shaping and coding lattices can lead to a more granular set of possible spectral efficiencies, which allows us to have more flexibility in choosing data rates as needed. In Fig.~\ref{Fig:shapinggain}, the shaping gain $g_{\text{s}}$ as a function of $\beta$ is presented for VCs with a cubic coding lattice and the considered shaping lattices in Table \ref{tab:gammas}. Larger markers show the spectral efficiencies for bit mapping; smaller markers show the finer-grained granularity of spectral efficiencies without considering bit mapping, where $m$ does not have to be a power of $2$. 

\section{Achievable Information Rates}
AIRs indicate the maximum amount of information that can be transmitted reliably over the underlying channel. It assumes a perfect channel code with an infinite blocklength, thus can be used as an upper bound of the performance of coded modulation \cite{alvarado18}. For very large VCs, the calculation of MI is challenging, since it requires the coordinates of all constellation points. In this section, we propose an MI estimation method and a log-likelihood (LLR) estimation method for very large constellations and apply these methods to VCs with a cubic coding lattice. The MI of our designed VCs is investigated, and their AIRs are compared with conventional scaled VCs when an LDPC code is applied. 
\subsection{MI estimation method for very large constellations}\label{subsec:MIest_method}
For a memoryless discrete channel, the MI between the equally probable transmitted symbols $\bX$ and the received noisy symbols $\bY$ can be written as
\begin{align}\label{eq:MI1}
    I(\bX;\bY)\triangleq \frac{1}{M}\sum_{i=1}^{M} \int_{\mathbb{C}^n}  f_{\bY|\bX}(\by|\bx_{i})\log \frac{f_{\bY|\bX}(\by|\bx_{i})}{f_{\bY}(\by)}d\by,
\end{align}
where $\bx_{i}$ for $i=1,\ldots,M$ form the constellation $\X$.
By applying Monte Carlo integration, the MI can be approximated as
\begin{align}
        I(\bX;\bY)&\approx \frac{1}{N_{\text{s}}} \sum_{i=1}^{N_{\text{s}}}\int_{\mathbb{C}^n}  f_{\bY|\bX}(\by|\bx_{i})\log \frac{f_{\bY|\bX}(\by|\bx_{i})}{f_{\bY}(\by)}d\by\\
        &\approx\frac{1}{N_{\text{s}}} \sum_{i=1}^{N_{\text{s}}} \log \frac{f_{\bY|\bX}(\by_{i}|\bx_{i})}{f_{\bY}(\by_{i})},\label{eq:MImontecarlo}
\end{align}
where $\bx_{i}$ for $i=1,2,\ldots,N_{\text{s}}$ are $N_{\text{s}}$ symbols drawn uniformly from $\X$, and given a certain $\bx_{i}$, $\by_{i}$ is drawn from the conditional distribution of the channel $f_{\bY|\bX}(\by_{i}|\bx_{i})$.

To numerically estimate \eqref{eq:MImontecarlo}, the key point is to calculate the marginal distribution of $\by$
\begin{align}
     f_{\bY}(\by) &= \frac{1}{M} \sum_{\bx\in \X} f_{\bY|\bX}(\by|\bx)\label{eq:fy}.
\end{align}
The exact calculation of $f_{\bY}(\by)$ using \eqref{eq:fy} requires storing all $M$ constellation points, which is infeasible when $M$ is very large. Furthermore, approximating $f_{\bY}(\by)$ by standard Monte-Carlo techniques is very inaccurate, because for most realistic channel laws $f_{\bY|\bX}(\by|\bx)$, only a tiny fraction of all constellation points $\bx$ contribute significantly to the sum in \eqref{eq:fy}. 

To approximate \eqref{eq:fy}, we propose a method based on \emph{importance sampling}, which is a weighted sampling method that oversamples from the important region we are interested in, thus making Monte Carlo feasible \cite[Ch.~9]{mcbook}. This concept has been used in the machine learning field to estimated the MI between the original data and the learned representation \cite{ML1}, but as far as we know never to estimate the MI of a communication channel. The classic importance sampling approach is to sample from a new importance distribution that is proportional to the product of the distribution of the 
integral variable and the integrand \cite[Eq.~(9.3)]{mcbook}, to resemble the true expectation. For example, to simulate \eqref{eq:fy} in our case, the importance distribution of $\bx \in \X$ should be proportional to $(1/M) f_{\bY|\bX}(\by|\bx)$. For the MI estimation problem, we refine this concept by using a step-wise sampling rule, in which we only sample symbols from an important set $\I(\by)$ which contains points that have an important contribution to $f_{\bY}(\by)$, and no samples are sampled from the complementary set $\X-\I(\by)$ since their contribution to $f_{\bY}(\by)$ is negligible compared with $\I(\by)$. The important set $\I(\by)$ is divided into $D$ disjoint subsets $\I_{d}(\by)$ for $d=1,\ldots,D$. The contribution to $f_{\bY}(\by)$ from each subset $\I_{d}(\by)$ can be either calculated by enumerating every symbol in $\I_{d}(\by)$, or estimated by performing a uniform Monte Carlo sampling in $\I_{d}(\by)$. Thus, the sum in \eqref{eq:fy} can be approximated as
\begin{align}
     f_{\bY}(\by) &\approx \frac{1}{M}\sum_{\bx \in \I(\by) } f_{\bY|\bX}(\by|\bx)\label{eq:estfy_C}\\
     &\approx \frac{1}{M}\sum_{d=1}^D \frac{|\I_d(\by)|}{K_d} \sum_{j=1}^{K_d} f_{\bY|\bX}(\by|\bx_{d,j}) \label{eq:estfy_Cr},
\end{align}
where $\bx_{d,j}$ for $j=1,\ldots,K_{d}$ are all points from $\I_{d}(\by)$ if ${K_{d}=|\I_d(\by)|}$, or $K_{d}$ uniform random samples from $\I_{d}(\by)$ if ${K_d < |\I_d(\by)|}$. For accurate results, the sampling sets $\I(\by)$ and $\I_{d}(\by)$, but not necessarily their sizes $K_{d}$, should be chosen as functions of $\by$.

We propose \eqref{eq:estfy_Cr} as a very general way to estimate $f_{\bY}(\by)$ and thereby the MI and related quantities. The proposed MI estimation method is a special case of importance sampling. Unlike the importance distribution in classical importance sampling, which should be more ``continuous'', our refined step-wise sampling rule makes random generation of $\bx$ easier. As the number of the elements in $\I(\by)$ increases, the estimated distribution $f_{\bY}(\by)$ in \eqref{eq:estfy_C} should converge to the exact value. As a special case, setting $\I(\by) = \X$ and $D=1$ in \eqref{eq:estfy_Cr} yields the exact expression \eqref{eq:fy} if $K_1 = M$ and a standard Monte Carlo estimate thereof if $K_1 < M$. The same idea can be applied to other structured constellations, or other similar problems. The readers can derive their own estimation rules from \eqref{eq:estfy_Cr} for a specific channel and constellation. 

\subsection{MI estimation for the designed VCs}\label{subsec:MIest_forVC}
We consider an $n$-dimensional real AWGN channel, which has the conditional distribution 
\begin{align}
    f_{\bY|\bX}(\by|\bx)=\frac{1}{(2\pi \sigma^2/n)^{n/2}}\exp{(-\frac{\left\| \by-\bx\right\|^2}{2\sigma^2/n})},
\end{align} 
where $\sigma^2$ is the total noise power. We define the signal-to-noise ratio (SNR) as $E_{\text{s}}/\sigma^2$.

For estimating the MI of our VCs with a cubic coding lattice, first, given a received noisy symbol $\by$, we define a Euclidean ball $\B(\by,R)$ containing all $n$-dimensional points in the translated cubic coding lattice having a Euclidean distance within $R$ from the nearest integer vector of $\by+\ba$, i.e.,
\begin{align}
    \B(\by,R)\triangleq\{\bx:\|\bx+\ba-\left \lfloor \by+\ba  \right \rceil\| \leq R,~\bx+\ba \in \Z^n\},
\end{align}
where the squared radius $R^2 \in \mathbb{N}$. The Euclidean ball $\B(\by,R)$ consists of $R^2+1$ Euclidean ``shells'', each of which contains all $n$-dimensional points in the translated cubic coding lattice having a Euclidean distance of $r$ from $\left \lfloor \by+\ba  \right \rceil$, i.e.,
\begin{align}
\mathcal{S}(\by,r)\triangleq \{\bx:  \|\bx+\ba-\left \lfloor \by+\ba \rceil \right \| =r, ~\bx +\ba \in \Z^n\},
\end{align}
for $r^2=0,1,\ldots,R^2$. The number of points in each shell $\mathcal{S}(\by,r)$ of a Euclidean ball $\B(\by,R)$ are listed in Table \ref{tab:numpointsinR}.

\begin{table}[tbp]
  \renewcommand{\arraystretch}{1.3}
  \caption{The number of points in each shell $\mathcal{S}(\by,r)$ of the Euclidean ball $\B(\by,R)$ ($n\geq 4 $ in this table).}
  \label{tab:numpointsinR}
  \centering
  \begin{tabular}{c c c c c c c}
    \hline 
    $r^2$ & $0$ & $1$ & $2$ & $3$ & $4$ & $\ldots$ \\
    \hline \hline
    $|\mathcal{S}(\by,r)|$ & $1$ & $2n$ & $2^2{n\choose2}$ & $2^3{n\choose3}$& $ 2^4{n\choose4} + 2{n\choose1}$& $\ldots$\\
    \hline
  \end{tabular}
\end{table}
The important set is defined as all points in $\mathcal{B}(\by,R)$ that belong to $\Gamma$ at the same time, i.e.,
\begin{align}
    \I(\by,R)=\B(\by,R) \cap \Gamma \label{eq:I},
\end{align}
which consists of $D=R^2+1$ disjoint subsets $\I_{d}(\by)$ for $d=1,\ldots,D$. Each subset $\I_{d}(\by)$ contains all points in $\mathcal{S}(\by,\sqrt{d-1})$ that also belong to $\Gamma$, i.e., ${\I_{d}(\by)=\mathcal{S}(\by,\sqrt{d-1}) \cap \Gamma}$.

How should $K_{d}$ and $D$ be chosen for accurate estimation? Heuristically, we have found that $K_d = \min\{|\I_d(\by)|, 10^4\}$ works well for all $d=1,\ldots,D$, which means that the contributions to \eqref{eq:estfy_Cr} from small enough subsets are computed exactly, whereas large subsets are sampled using $10^4$ random points. To determine $D$, we evaluate \eqref{eq:estfy_Cr} for increasing values of $D$ until the relative increase is less than $0.5\%$. More precisely, denoting the estimated $f_{\bY}(\by)$ using \eqref{eq:estfy_Cr} for a certain $D$ by $f_{\bY}^{(D)}(\by)$, we choose the smallest integer $D$ for which 
\begin{align}
\max\left(  \frac{f_{\bY}^{(D+1)}(\by)-f_{\bY}^{(D)}(\by)}{f_{\bY}^{(D)}(\by)} \right) < 0.5\%,\label{eq:criteria}
\end{align} 
where the maximum is taken over multiple random vectors $\by$ and multiple Monte-Carlo realizations of \eqref{eq:estfy_Cr}. If a range of SNRs is being considered, we usually apply the criterion \eqref{eq:criteria} to the lowest SNR, which intuitively needs the largest number of subsets, and then use the obtained value of $D$ throughout the SNR range. Alternatively, \eqref{eq:criteria} can be evaluated separately for each SNR to save complexity. With these parameters, \eqref{eq:estfy_Cr} is able to estimate $f_{\bY}(\by)$ accurately for VCs with up to $M=16777216$ points for all SNRs. For larger constellations where we have no benchmarks to compare with, we conjecture that this method is still valid. After having reliable estimated values of $f_{\bY}(\by)$, the MI can be estimated accurately by \eqref{eq:MImontecarlo}.

\begin{figure}[tbp]
\centering
\begin{tikzpicture}

	\begin{semilogyaxis}[
		xmin=1,
		xmax=21,
		xticklabels={$1/1$,$5/89$,$9/321$,$13/761$,$17/1281$,$21/4785$},
		xtick={1,5,9,13,17,21},
		ymin=5e-8, ymax=1e-3,
		xlabel={$D$ and $|\B(\by,\sqrt{D-1})|$ },
		ylabel={$pf_{\bY}^{(D)}(\by)$},
		ylabel style={at={(axis description cs:-0.01,0.5)}, anchor=north},
		cycle list name=myCycleList,
	    legend pos=north west,
		legend cell align=left,
		legend style={fill=white},
		ylabel style={yshift=.2cm},
		xlabel style={xshift=-.05cm},
		height =0.4\textwidth,
		width=0.48\textwidth,
	]
	\addplot+[lines-1] table[
		x=radius,
		y=snr14,
	] {./figures/fy/estfy_D4.txt};
	\addplot+[lines-2] table[
		x=radius,
		y=snr16,
	] {./figures/fy/estfy_D4.txt};
	\addplot+[lines-3] table[
		x=radius,
		y=snr18,
	] {./figures/fy/estfy_D4.txt};
	\addplot+[lines-4] table[
		x=radius,
		y=snr20,
	] {./figures/fy/estfy_D4.txt};
	\addplot+[lines-5] table[
		x=radius,
		y=snr22,
	] {./figures/fy/estfy_D4.txt};
	\addplot+[lines-6] table[
		x=radius,
		y=snr24,
	] {./figures/fy/estfy_D4.txt};
	\addplot+[lines-7] table[
		x=radius,
		y=snr26,
	] {./figures/fy/estfy_D4.txt};
	\addplot+[lines-8] table[
		x=radius,
		y=snr28,
	] {./figures/fy/estfy_D4.txt};
  	 \addplot+[dashed,color=black,mark=none]  coordinates {(0,0.00036156) (20,0.00036156)};
  	 \addplot+[dashed,color=black,mark=none] coordinates {(0,0.00014394) (20,0.00014394)};
  	 \addplot+[dashed,color=black,mark=none] coordinates {(0,3.9156e-05) (20,3.9156e-05)};
  	 \addplot+[dashed,color=black,mark=none] coordinates {(0,2.0295e-05) (20,2.0295e-05)};
  	 \addplot+[dashed,color=black,mark=none]  coordinates {(0,9.2801e-06) (20,9.2801e-06)};
  	 \addplot+[dashed,color=black,mark=none] coordinates {(0,3.0845e-06) (20,3.0845e-06)};
  	 \addplot+[dashed,color=black,mark=none] coordinates {(0,5.871e-07) (20,5.871e-07)};
  	 \addplot+[dashed,color=black,mark=none] coordinates {(0,7.0641e-08) (20,7.0641e-08)};
  	  \node[shape=rectangle] (a) at (18.5, 5e-4) {\footnotesize{SNR $=14\;$dB}};
  	  \node[shape=rectangle] (a) at (18.5, 2e-4) {\footnotesize{SNR $=16\;$dB}};
  	  \node[shape=rectangle] (a) at (18.5, 5.5e-5) {\footnotesize{SNR $=18\;$dB}};
  	  \node[shape=rectangle] (a) at (18.5, 2.9e-5) {\footnotesize{SNR $=20\;$dB}};
  	  \node[shape=rectangle] (a) at (18.5, 1.4e-5) {\footnotesize{SNR $=22\;$dB}};
  	  \node[shape=rectangle] (a) at (18.5, 4.2e-6) {\footnotesize{SNR $=24\;$dB}};
  	  \node[shape=rectangle] (a) at (18.5, 8e-7) {\footnotesize{SNR $=26\;$dB}};
  	  \node[shape=rectangle] (a) at (18.5, 1e-7) {\footnotesize{SNR $=28\;$dB}};
 	\end{semilogyaxis}
\end{tikzpicture}
\caption{The estimated value $pf_{\bY}^{(D)}(\by)$ as a function of $D$ for different SNRs (solid curves with markers). The black dashed lines are the corresponding benchmark values $pf_{\bY}(\by)$. The number after the `/' is the corresponding $|\B(\by,\sqrt{D-1})|$. }
\label{Fig:fyD4}
\end{figure}
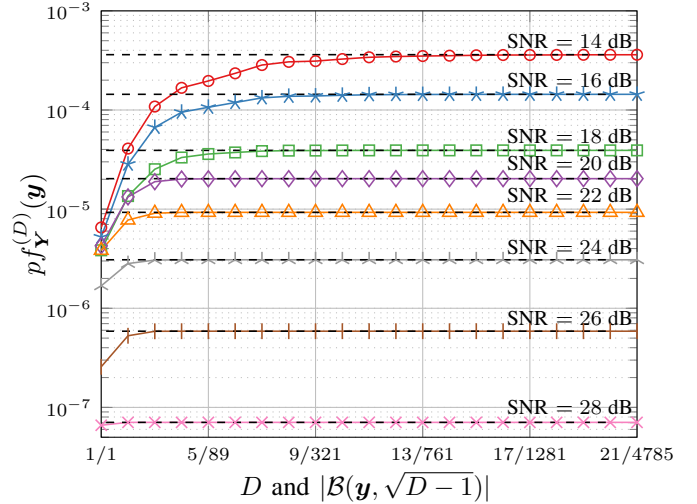

\emph{Example 1:} We consider a moderate-size VC based on the lattice partition $\Z^4/16D_{4}$ with $M=131072$ constellation points. For better visualization, we show $pf_{\bY}^{(D)}(\by)$ as a function of $D$ upon receiving a noisy symbol $\by$ in Fig. \ref{Fig:fyD4}, where $p=(2\pi \sigma^2/n)^{n/2}$ is a constant for a given $\sigma^2$. The benchmark values $pf_{\bY}(\by)$ are calculated using \eqref{eq:fy}. As $D$ increases, the estimated values all converge to the exact values for different SNRs in Fig. \ref{Fig:fyD4}. 

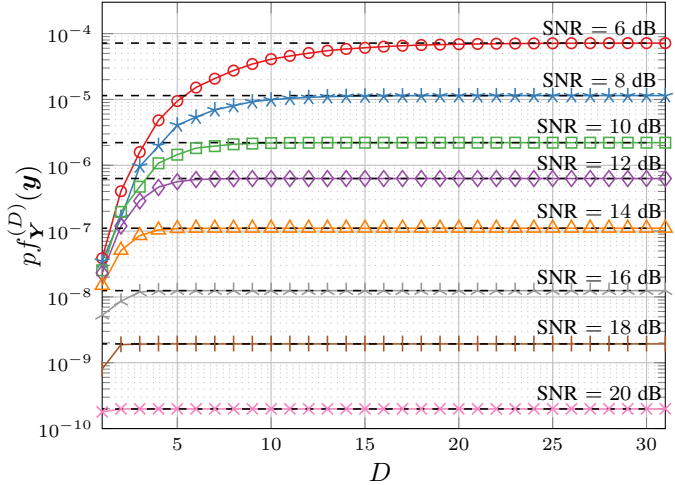
\begin{figure}[tbp]
\centering
\begin{tikzpicture}

	\begin{semilogyaxis}[
		xmin=1,
		xmax=31,
		ymin=1e-10, ymax=3e-4,
		xlabel={$D$},
		ylabel={$pf_{\bY}^{(D)}(\by)$},
		ylabel style={at={(axis description cs:-0.02,0.5)}, anchor=north},
		cycle list name=myCycleList,
	    legend pos=north west,
		legend cell align=left,
		legend style={fill=white},
		ylabel style={yshift=.2cm},
		xlabel style={xshift=-.05cm},
		height =0.4\textwidth,
		width=0.5\textwidth,
	]
	\addplot+[lines-1,mark=none] table[
		x=radius,
		y=snr6,
	] {./figures/fy/estfy_E8.txt};
	\addplot+[lines-2,mark=none] table[
		x=radius,
		y=snr8,
	] {./figures/fy/estfy_E8.txt};
	\addplot+[lines-3,mark=none,mark=none] table[
		x=radius,
		y=snr10,
	] {./figures/fy/estfy_E8.txt};
	\addplot+[lines-4,mark=none] table[
		x=radius,
		y=snr12,
	] {./figures/fy/estfy_E8.txt};
	\addplot+[lines-5,mark=none] table[
		x=radius,
		y=snr14,
	] {./figures/fy/estfy_E8.txt};
	\addplot+[lines-6,mark=none] table[
		x=radius,
		y=snr16,
	] {./figures/fy/estfy_E8.txt};
	\addplot+[lines-7,mark=none] table[
		x=radius,
		y=snr18,
	] {./figures/fy/estfy_E8.txt};
	\addplot+[lines-8,mark=none] table[
		x=radius,
		y=snr20,
	] {./figures/fy/estfy_E8.txt};
	
	\addplot+[only marks,color=lines-1] table[
		x=radius,
		y=snr6,
	] {./figures/fy/estfy_sampled_E8.txt};
	\addplot+[only marks,color=lines-2] table[
		x=radius,
		y=snr8,
	] {./figures/fy/estfy_sampled_E8.txt};
	\addplot+[only marks,color=lines-3] table[
		x=radius,
		y=snr10,
	] {./figures/fy/estfy_sampled_E8.txt};
	\addplot+[only marks,color=lines-4] table[
		x=radius,
		y=snr12,
	] {./figures/fy/estfy_sampled_E8.txt};	
	\addplot+[only marks,color=lines-5] table[
		x=radius,
		y=snr14,
	] {./figures/fy/estfy_sampled_E8.txt};
	\addplot+[only marks,color=lines-6] table[
		x=radius,
		y=snr16,
	] {./figures/fy/estfy_sampled_E8.txt};	
	\addplot+[only marks,color=lines-7] table[
		x=radius,
		y=snr18,
	] {./figures/fy/estfy_sampled_E8.txt};
	\addplot+[only marks,color=lines-8] table[
		x=radius,
		y=snr20,
	] {./figures/fy/estfy_sampled_E8.txt};
	
  	 \addplot+[dashed,color=black,mark=none]  coordinates {(0,7.1861e-05) (30,7.1861e-05)};
  	 \addplot+[dashed,color=black,mark=none] coordinates {(0,1.1446e-05) (30,1.1446e-05)};
  	 \addplot+[dashed,color=black,mark=none] coordinates {(0,2.2074e-06) (30,2.2074e-06)};
  	 \addplot+[dashed,color=black,mark=none] coordinates {(0,6.2954e-07) (30,6.2954e-07)};
  	 \addplot+[dashed,color=black,mark=none]  coordinates {(0,1.105e-07) (30,1.105e-07)};
  	 \addplot+[dashed,color=black,mark=none] coordinates {(0,1.2456e-08) (30,1.2456e-08)};
  	 \addplot+[dashed,color=black,mark=none] coordinates {(0,1.9325e-09) (30,1.9325e-09)};
  	 \addplot+[dashed,color=black,mark=none] coordinates {(0,1.986e-10) (30,1.986e-10)};
  	  \node[shape=rectangle] (a) at (27.5, 1.2e-4) {\footnotesize{SNR $=6\;$dB}};
  	  \node[shape=rectangle] (a) at (27.5, 2e-5) {\footnotesize{SNR $=8\;$dB}};
  	  \node[shape=rectangle] (a) at (27.5, 4e-6) {\footnotesize{SNR $=10\;$dB}};
  	  \node[shape=rectangle] (a) at (27.5, 1.1e-6) {\footnotesize{SNR $=12\;$dB}};
  	  \node[shape=rectangle] (a) at (27.5, 2e-7) {\footnotesize{SNR $=14\;$dB}};
  	  \node[shape=rectangle] (a) at (27.5, 2e-8) {\footnotesize{SNR $=16\;$dB}};
  	  \node[shape=rectangle] (a) at (27.5, 3.5e-9) {\footnotesize{SNR $=18\;$dB}};
  	  \node[shape=rectangle] (a) at (27.5, 3.5e-10) {\footnotesize{SNR $=20\;$dB}};
 	\end{semilogyaxis}
\end{tikzpicture}
\caption{The estimated value $pf_{\bY}^{(D)}(\by)$ as a function of $D$ for different SNRs. Solid lines without markers are estimated with $K_{d}=|\I_d(\by)|$ for all subsets. The markers are estimated with $K_{d}=10^4$ uniform samples from $\I_d(\by)$ for subsets with $d>8$ and $K_{r}=|\I_d(\by)|$ for subsets with $1 \leq d\leq 8$. The black dashed lines are the corresponding benchmark values $pf_{\bY}(\by)$.}
\label{Fig:fyE8}
\end{figure}

\emph{Example 2:} We consider a larger-size VC of the lattice partition $\Z^8/8E_{8}$ with $M=16777216$ constellation points. Table \ref{tab:r2E8} lists the minimum required $D$ found by \eqref{eq:criteria} and the corresponding number of symbols in $\B(\by,\sqrt{D-1}))$ at different SNRs, which shows that $|\B(\by,\sqrt{D-1})|$ grows fast as the SNR decreases. Fig. \ref{Fig:fyE8} shows the convergence of $pf_{\bY}^{(D)}(\by)$ to $pf_{\bY}(\by)$ at medium SNR range. We observe from Fig. \ref{Fig:fyE8} that the chosen values of $D$ in Table~\ref{tab:r2E8} are sufficient to approach $f_{\bY}(\by)$ correctly for these SNR values. 

\begin{table}
  \setlength{\tabcolsep}{4.5pt}
  \renewcommand{\arraystretch}{1.3}
  \caption{The minimum required $D$ and number of points in $\B(\by,\sqrt{D-1})$ for the convergence of $f_{\bY}^{(D)}(\by)$ to $f_{\bY}(\by)$ for the VC of the lattice partition $\Z^8/8E_{8}$ at different SNRs.}
  \label{tab:r2E8}
  \centering
  \begin{tabular}{c c c c c c c}
    \hline 
    SNR & $16$ & $14$ & $12$ & $10$ & $8$ &$6$\\
    \hline \hline
    $D$  & $5$ & $6$ & $8$ & $11$ &$16$& $23$\\
    \hline
    $|\B(8,\sqrt{D-1})|$ &$1713$&  $3729$& $12369$& $47921$& $231185$&$1025649$ \\
    \hline
  \end{tabular}
\end{table}

\subsection{LLR approximation for very large constellations}
We propose an LLR approximation method for very large constellations and exemplify it for our designed VCs and the scaled VCs.

For transmission of a constellation $\X$ through a given channel with conditional distribution $f_{\bY|\bX}(\by|\bx)$, upon receiving a noisy symbol $\by$, the LLR of a certain bit $b_{i}$, for $i=1,2,\ldots,\log_{2}(|\X|)$, is defined as
\begin{align}
    \text{LLR}(b_{i}|\by)&\triangleq\log{\frac{\text{Pr}(b_{i}=0|\by)}{\text{Pr}(b_{i}=1|\by)}}\notag\\
    &=\log \frac{\sum_{\bx\in \X^{(i,0)}}f_{\bY|\bX}(\by|\bx)}{\sum_{\bx\in \X^{(i,1)}}f_{\bY|\bX}(\by|\bx)}\label{eq:exactLLR},
\end{align}
where $\X^{(i,0)}$ and $\X^{(i,1)}$ are the sets of constellation points with $0$ and $1$ at position $i$, respectively. The exact LLR calculated using the whole constellation is accurate but too complex to compute, which can be approximated by only considering the most likely constellation point with $b_{i}=0$ (or $b_{i}=1$), i.e.,
\begin{align}
    \text{LLR}(b_{i}|\by)\approx \log \frac{\max_{\bx\in \X^{(i,0)}}f_{\bY|\bX}(\by|\bx)}{\max_{\bx\in \X^{(i,1)}}f_{\bY|\bX}(\by|\bx)}.\label{eq:LLR}
\end{align}
However, for very large constellations, searching for the most likely constellation point with $b_{i}=0$ and $b_{i}=1$ from the whole constellation is still infeasible. 

We propose a similar technique as our MI estimation method in \ref{subsec:MIest_method} to further approximate the LLR. Instead of searching points from $\X$, we only search the closest point from an important set $\I(\by)$. Specifically, the LLR in \eqref{eq:LLR} is further approximated as
\begin{align}
    \text{LLR}(b_{i}|\by)\approx \log \frac{\max_{\bx\in \I^{(i,0)}(\by)}f_{\bY|\bX}(\by|\bx)}{\max_{\bx\in \I^{(i,1)}(\by)}f_{\bY|\bX}(\by|\bx)},\label{eq:approxLLR}
\end{align}
where $\I^{(i,0)}(\by) = \X^{(i,0)} \cap \I(\by)$ and $\I^{(i,1)}(\by) = \X^{(i,1)} \cap \I(\by)$. If there is no constellation point in $\I^{(i,0)}(\by)$ (or $\I^{(i,1)}(\by)$), we assume the most likely constellation point has a small probability, i.e., setting $\max_{\bx\in \I^{(i,0)}(\by)}f_{\bY|\bX}(\by|\bx)$ (or $\max_{\bx\in \I^{(i,1)}(\by)}f_{\bY|\bX}(\by|\bx)$) to a small 
default value. 

For the AWGN channel, the widely-used approximate LLR derived from \eqref{eq:LLR} is
\begin{align}
    &\text{LLR}(b_{i}|\by)\approx \notag\\ &-\frac{1}{2\sigma^2/n}\left (\min_{\bx\in\X^{(i,0)}}(\|\by-\bx\|^2)- \min_{\bx\in\X^{(i,1)}}(\|\by-\bx\|^2)\right),\label{eq:LLR_awgn}
\end{align}
where only the nearest constellation point with $b_{i}=0$ (or $b_{i}=1$) to $\by$ are considered \cite[Eq.~(6)]{viterbi98}. Analogously with \eqref{eq:approxLLR}, we further approximate \eqref{eq:LLR_awgn} as
\begin{align}
    &\text{LLR}(b_{i}|\by)\approx \notag \\ &-\frac{1}{2\sigma^2/n}\left (\min_{\bx\in\I^{(i,0)}(\by)}(\|\by-\bx\|^2)- \min_{\bx\in\I^{(i,1)}(\by)}(\|\by-\bx\|^2)\right),\label{eq:LLR_awgn_approx}
\end{align}

For our designed VCs with a cubic coding lattice, the important set is defined as in \eqref{eq:I} with a radius $R$. This parameter provides a trade-off between low computational complexity (small $R$) and good decoding performance (high $R$). If ${\I^{(i,0)}(\by)=\varnothing}$, then setting ${\max_{\bx\in \I^{(i,0)}(\by)}f_{\bY|\bX}(\by|\bx)}$ to a small probability is equivalent to setting ${\min_{\bx\in \I^{(i,0)}(\by)}\|\by-\bx\|^2}$ to a large default value $q$ which is larger than $R^2$, and the same rule applies to $\I^{(i,1)}(\by)$. A default value close to the boundary of the important region (e.g., $q=R^2+1$) is usually not a good choice, neither a very large $q$ (e.g., $q=100 R^2$). The decoding performance can be roughly optimized by testing different $q$ for a given $R$.

Similarly, for very large scaled VCs, the important set can be generalized as
\begin{align}
    \mathcal{I}(\by)\triangleq \{\bx:\|\bx+\ba-\Q_{\Lambda}(\by+\ba)\| \leq R,\bx \in \Gamma\}.
\end{align}
Then \eqref{eq:approxLLR} can be used to estimate the LLR for the scaled VCs.
\subsection{Results}
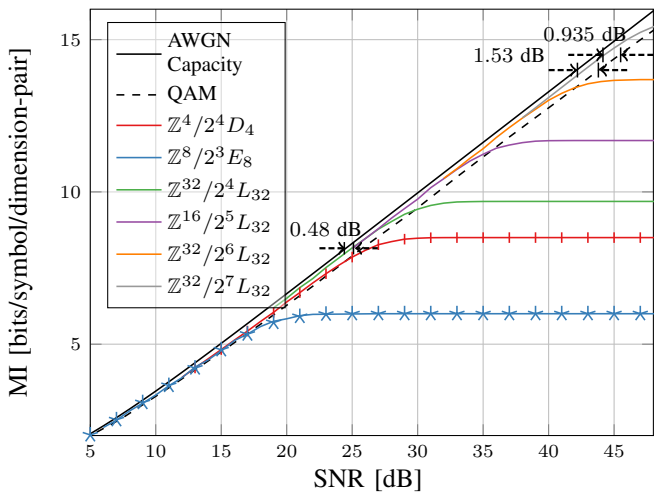
\begin{figure}[tbp]
\centering
\begin{tikzpicture}

	\begin{axis}[
		xmin=5,
		xmax=48,
		ymin=2, ymax=16,
		xlabel={SNR [dB]},
		ylabel={MI [bits/symbol/dimension-pair]},
		ylabel style={at={(axis description cs:-0.01,0.5)}, anchor=north},
		cycle list name=myCycleList,
	    legend pos=north west,
		legend cell align=left,
		legend style={fill=white,fill opacity=0.4, draw opacity=1,text opacity=1},
		ylabel style={yshift=.1cm},
		xlabel style={xshift=-.05cm},
		height =0.4\textwidth,
		width=0.5\textwidth,
	]
	\addplot+[black,mark=none] table[
		x=SNR,
		y=C,
	] {./figures/MI/Capacity.txt};\addlegendentry{AWGN\\Capacity}
	\addplot+[black, dashed ,mark=none] table[
		x=SNR,
		y=I,
	] {./figures/MI/I_qam.txt};\addlegendentry{QAM}
	\addplot+[lines-1,mark=none] table[
		x=SNR,
		y=I,
	] {./figures/MI/I_D4.txt};\addlegendentry{$\Z^4/2^4D_{4}$}

	\addplot+[lines-2,mark=none] table[
		x=SNR,
		y=I,
	] {./figures/MI/I_E8.txt};\addlegendentry{$\Z^8/2^3E_{8}$}
	\addplot+[lines-3,mark=none] table[
		x=SNR,
		y=I,
	] {./figures/MI/I_L32_r4.txt};\addlegendentry{$\Z^{32}/2^4L_{32}$}
	\addplot+[lines-4,mark=none] table[
		x=SNR,
		y=I,
	] {./figures/MI/I_L32_r5.txt};\addlegendentry{$\Z^{16}/2^5L_{32}$}
	\addplot+[lines-5,mark=none] table[
		x=SNR,
		y=I,
	] {./figures/MI/I_L32_r6.txt};\addlegendentry{$\Z^{32}/2^6L_{32}$}
	\addplot+[lines-6,mark=none] table[
		x=SNR,
		y=I,
	] {./figures/MI/I_L32_r7.txt};\addlegendentry{$\Z^{32}/2^7L_{32}$}
	\addplot+[only marks,color=lines-1,mark=|] table[
		x=SNR,
		y=I,
	] {./figures/MI/Ireal_D4.txt};
	\addplot+[only marks,color=lines-2] table[
		x=SNR,
		y=I,
	] {./figures/MI/Ireal_E8.txt};
    \node[anchor=east] (source1) at (axis cs:40,14){};
      \node[] (destination1) at (axis cs:42.9,14){};
      \node[] (top1) at (axis cs:42.2,14.55){};
      \node[] (bot1) at (axis cs:42.2,13.45){};
      \draw[line width=0.2mm,-](top1)--(bot1);
      \draw[line width=0.3mm,dashed,dash pattern=on 0.8mm off 0.4mm,->](source1)--(destination1);
       
    \node[anchor=west] (source2) at (axis cs:46,14){};
      \node[] (destination2) at (axis cs:43.2,14){};
      \node[] (top2) at (axis cs:43.8,14.55){};
      \node[] (bot2) at (axis cs:43.8,13.45){};
      \draw[line width=0.2mm,-](top2)--(bot2);
       \draw[line width=0.3mm,dashed,dash pattern=on 0.8mm off 0.4mm,->](source2)--(destination2);
    \node[shape=rectangle] (b) at (37,14.5) {\footnotesize{$1.53$ dB}};
    \node[anchor=east] (source3) at (axis cs:22.5,8.15){};
      \node[] (destination3) at (axis cs:25.2,8.15){};
     \node[] (top3) at (axis cs:24.4,8.66){};
      \node[] (bot3) at (axis cs:24.4,7.64){};
      \draw[line width=0.2mm,-](top3)--(bot3);
       \draw[line width=0.3mm,dashed,dash pattern=on 0.8mm off 0.4mm,->](source3)--(destination3);
    \node[anchor=west] (source4) at (axis cs:27,8.15){};
      \node[] (destination4) at (axis cs:24.5,8.15){};
      \node[] (top4) at (axis cs:25.1,8.66){};
      \node[] (bot4) at (axis cs:25.1,7.64){};
      \draw[line width=0.2mm,-](top4)--(bot4);
       \draw[line width=0.3mm,dashed,dash pattern=on 0.8mm off 0.4mm,->](source4)--(destination4);
    \node[shape=rectangle] (b) at (23,8.8) {\footnotesize{$0.48$ dB}}{};
    
    \node[anchor=east] (source5) at (axis cs:41.5,14.5){};
      \node[] (destination5) at (axis cs:44.9,14.5){};
     \node[] (top5) at (axis cs:44.15,15.05){};
      \node[] (bot5) at (axis cs:44.15,13.95){};
      \draw[line width=0.2mm,-](top5)--(bot5);
       \draw[line width=0.3mm,dashed,dash pattern=on 0.8mm off 0.4mm,->](source5)--(destination5);
    \node[anchor=west] (source6) at (axis cs:48,14.5){};
      \node[] (destination6) at (axis cs:44.8,14.5){};
      \node[] (top6) at (axis cs:45.5,15.05){};
      \node[] (bot6) at (axis cs:45.5,13.95){};
      \draw[line width=0.2mm,-](top6)--(bot6);
       \draw[line width=0.3mm,dashed,dash pattern=on 0.8mm off 0.4mm,->](source6)--(destination6);
    \node[shape=rectangle] (b) at (42.8,15.2) {\footnotesize{$0.935$ dB}}{};
 	\end{axis}
\end{tikzpicture}
\caption{The estimated MI as a function of the SNR for multidimensional VCs (solid curves without markers). The markers are the MI estimated using the exact $f_{\bY}(\by)$ by \eqref{eq:fy}.}
\label{Fig:MI}
\end{figure}

 Fig. \ref{Fig:MI} shows the estimated MI for multidimensional VCs using our proposed estimation method in \ref{subsec:MIest_method} and parameters are chosen as suggested in \ref{subsec:MIest_forVC}. To validate the correctness of our estimation, the MI simulated with the exact $f_{\bY}(\by)$ in \eqref{eq:fy} for some moderate-size VCs are included as benchmarks. It shows that our estimated MI is consistent with the benchmark MIs (markers). With $L_{32}$ as the shaping lattice, at SNR $=45$ dB, we can observe the maximum $0.935$ dB shaping gain. At SNR $=25$ dB, the gap to capacity is reduced from $1.33$ dB to $0.48$ dB compared with the QAM constellation.
 
\begin{figure}[htbp]
\centering
\begin{tikzpicture}

	\begin{axis}[
		xmin=15,
		xmax=50,
		ymin=0, ymax=14,
		xlabel={SNR [dB]},
		ylabel={AIR [bits/symbol/dimension-pair]},
		ylabel style={at={(axis description cs:-0.01,0.5)}, anchor=north},
		cycle list name=myCycleList,
	    legend pos=south east,
		legend cell align=left,
		legend style={fill=white, fill opacity=0.4, draw opacity=1,text opacity=1},
		ylabel style={yshift=.1cm},
		xlabel style={xshift=-.05cm},
		height =0.4\textwidth,
		width=0.5\textwidth,
	]
	\addplot+[black,mark=none] table[
		x=SNR,
		y=C,
	] {./figures/MI/Capacity.txt};\addlegendentry{AWGN Capacity}
	\addplot+[black,dashed,mark=none] table[
		x=SNR,
		y=I,
	] {./figures/MI/I_qam.txt};\addlegendentry{MI QAM}
	\addplot+[lines-1] table[
		x=SNR,
		y=I,
	] {./figures/LDPC/Icoded_D4_cubic_whole.txt};\addlegendentry{LDPC $\Z^4/2^4D_{4}$}
	\addplot+[dashed,lines-1,mark=star,mark options=solid] table[
		x=SNR,
		y=I,
	] {./figures/LDPC/Icoded_D4_fine_whole.txt};\addlegendentry{LDPC $D_{4}/2^4D_{4}$}
	\addplot+[lines-2] table[
		x=SNR,
		y=I,
	] {./figures/LDPC/Icoded_D4_cubic.txt};\addlegendentry{LDPC $\Z^4/2^6D_{4}$}
	\addplot+[lines-3] table[
		x=SNR,
		y=I,
	] {./figures/LDPC/Icoded_E8_cubic.txt};\addlegendentry{LDPC $\Z^8/2^6E_{8}$}
	\addplot+[lines-4,mark=diamond] table[
		x=SNR,
		y=I,
	] {./figures/LDPC/Icoded_L16_cubic.txt};\addlegendentry{LDPC $\Z^{16}/2^6\Lambda_{16}$}
	\addplot+[lines-5] table[
		x=SNR,
		y=I,
	] {./figures/LDPC/Icoded_L32_cubic.txt};\addlegendentry{LDPC $\Z^{32}/2^6L_{32}$}

 	\end{axis}
\end{tikzpicture}
\caption{The estimated AIR as a function of the SNR for multidimensional VCs with the DVB-S2 LDPC codes. The code rate $R_{\text{c}}\in \{1/3, 1/2, 3/5, 2/3, 3/4, 5/6, 9/10\}$. }
\label{Fig:LDPC}
\end{figure}
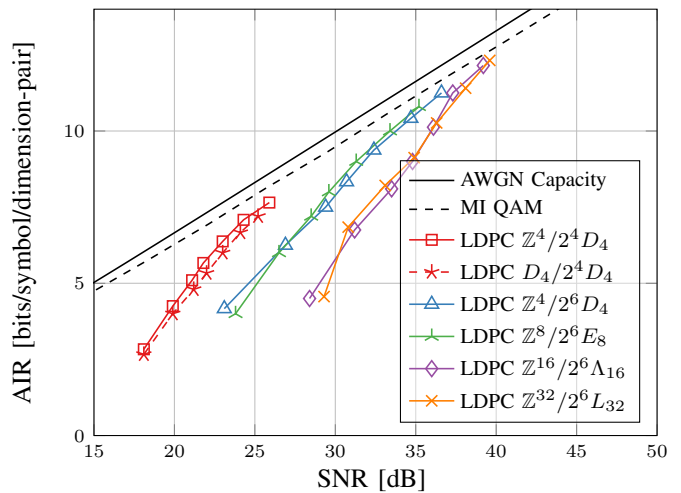

 We also investigate the performance of our VCs in coded systems. An LDPC code\footnote{The codeword length is $64800$ and $50$ decoding iterations are used.} from the digital video broadcasting (DVB-S2) standard \cite{dvbs2} with multiple code rates are applied to our designed VCs. Our proposed LLR approximation in \eqref{eq:LLR_awgn_approx} is applied. Fig.~\ref{Fig:LDPC} shows the estimated required SNRs to achieve a BER below $10^{-4}$ after LDPC decoding for the VCs based on the lattice partition $\Z^4/2^6D_{4}$ ($R^2=20,\;q=50$), $\Z^8/2^6E_{8}$ ($R^2=6,\;q=20$), $\Z^{16}/2^6\Lambda_{16}$ ($R=3,\;q=20$), and $\Z^{32}/2^6L_{32}$ ($R^2=2,\;q=13$). The default values $q$ are not globally optimized for different VCs, since our goal is not to design an LDPC code to maximize the AIRs. For comparison, we present results for the scaled VC based on the lattice partition $D_{4}/2^4D_{4}$ and our VC based on the lattice partition $\Z^{4}/2^4D_{4}$ simulated using the LLR in \eqref{eq:LLR_awgn}, i.e., without the further approximation in \eqref{eq:LLR_awgn_approx}. The result shows that our VC with a cubic coding lattice can have slightly higher AIRs than the scaled VC. In this case, the loss of coding gain due to the cubic coding lattice is more than compensated by the usage of a lower-rate LDPC code to obtain the same net AIR, which is consistent with the BER improvements observed in \cite[Fig.~4]{ourISIT}. The better AIR performance might come from that a pseudo-Gray labeling is more efficient for the cubic coding lattice, since our VC $\Z^{4}/2^4D_{4}$ has a Gray penalty of $1.01$, which is half of that for the scaled VC $D_{4}/2^4D_{4}$ ($1.99$).

\section{Conclusion}
We proposed Voronoi constellations with a cubic coding lattice, for which we follow Kurkoski's encoding and decoding algorithms and apply pseudo-Gray labeling to minimize the BER. As a structured geometric shaping method, this class of VC has high shaping gains of up to $1.03$ dB and low complexity. Thanks to its cubic coding lattice, an MI estimation method for very large constellation size based on importance sampling is proposed for the first time, which enables us to observe an up to $0.85$ dB shaping gain for medium SNR values. Our proposed LLR approximation method makes the analysis of AIRs of our VCs in coded systems at high spectral efficiencies possible, which is infeasible for the scaled VCs. Compared with the conventional scaled VCs, our designed VCs have the advantages that 1) the decoding algorithm is simpler, 2) the spectral efficiency for mapping integers to bits is improved to $1$ [bit/symbol/dimension-pair] realized by rotating and scaling the shaping lattices, 3) the AIRs after combining with an LDPC code can be higher due to our better pseudo-Gray labeling, and 4) analysing the MI and AIR becomes feasible for very large constellation sizes.


%




\ifCLASSOPTIONcaptionsoff
  \newpage
\fi



\bibliographystyle{IEEEtran}
\bibliography{main.bib}
\balance
\end{document}